\documentclass[twocolumn,floatfix,showpacs,superscriptaddress]{revtex4-1} %

\usepackage{mathptmx} 
\usepackage{soul}

\usepackage{amsmath}
\usepackage{graphicx,epstopdf}
\usepackage[utf8]{inputenc}

\newcommand{\ket}[1]{\ensuremath{\left|#1\right\rangle}}

\newcommand{\expval}[1]{\ensuremath{\langle #1 \rangle}}

\newcommand{\abs}[1]{\ensuremath{\left| #1 \right|}}
\newcommand{\av}[1]{\ensuremath{\left\langle #1 \right\rangle}}

\newcommand{\vk}{\mathbf{k}}
\newcommand{\vp}{\mathbf{p}}
\newcommand{\vq}{\mathbf{q}}

\newcommand{\vR}{\mathbf{R}}

\begin{document}

\title{Novel pairing mechanism for superconductivity at a vanishing level of doping driven by critical ferroelectric modes}

\author{Yaron Kedem}
\affiliation{Department of Physics, Stockholm University, AlbaNova University Center, 106 91 Stockholm, Sweden}

\begin{abstract}
Superconductivity (SC) occurring at low densities of mobile electrons is still a mystery since the standard theories do not apply in this regime. We address this problem by using a microscopic model for ferroelectric (FE) modes, which mediate an effective attraction between electrons. When the dispersion of modes, around zero momentum, is steep,  forward scattering is the main pairing process and the self-consistent equation for the gap function can be solved analytically. The solutions exhibit unique features: Different momentum components of the gap function are decoupled, and at the critical regime of the FE modes, different frequency components are also decoupled. This leads to effects that can be observed experimentally: The gap function can be non-monotonic in temperature and the critical temperature can be independent of the chemical potential. The model is applicable to lightly doped polar semiconductors, in particular, strontium titanate.
\end{abstract}

\maketitle

\emph{Introduction.}
Superconductivity is one of the most striking quantum phenomena in many body physics. More than a century after its discovery, it can be described by microscopic models only in a limited range of systems. The most prominent theory is BCS \cite{bcs}, which assumes an attractive interaction between electrons of the form
\begin{equation} \label{inter}
V(\vk,\vk') = \left\{ \begin{tabular}{ l r r}
  g, & ~~~&$ \xi_\vk,\xi_{\vk'} < \omega_D, $\\
  0, & ~~~&{ otherwise}, \\
\end{tabular} \right.
\end{equation}
where $\vk$ and $\vk'$ are the momenta of the electrons, $\xi_\vk$ and $\xi_{\vk'}$ are their energies, $g$ is a coupling constant and $\omega_D $ is the Debye frequency of the phonons mediating the interaction. The theory predicts a superconducting gap $\Delta =  2\omega_D e^{-1/ g N(0)}$, where $N(0)$ is the density of states at the Fermi surface and a transition temperature  $T_c \simeq 0.57 \Delta$. The form of the interaction in Eq. (\ref{inter}) can be reasonable in the so-called adiabatic regime $\omega_D \ll E_f$, where $E_f$ is the Fermi energy of the electrons, in which the ions responds slowly compare to the velocity of electrons. The retardation effect can result in an effective attraction and also goes hand in hand with a significant technical simplification, separating the dynamics of the electrons and those of the phonons so $N(0)$ is the only relevant quantity regarding the electrons. In the non-adiabatic regime it is hard to imagine that  Eq. (\ref{inter}) is applicable and a different physical picture is required. 

\begin{figure}
\centering
\includegraphics[trim=4.1cm 4.7cm 7.5cm 3.3cm, clip=true,width=0.49\textwidth]{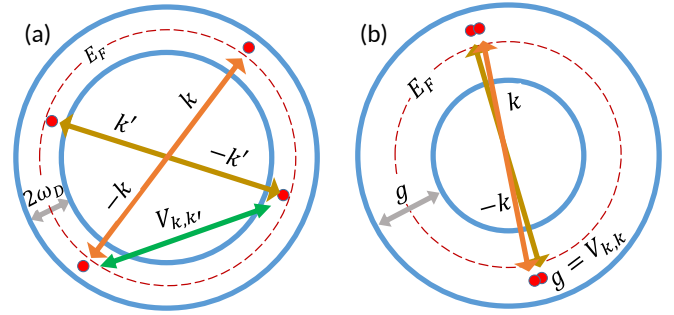}
\caption{An illustration of the pairing mechanism. (a) The standard BCS picture: Cooper pairs of quasi-particles with opposite momentum interact only when their energy is within $\omega_D$ from the Fermi surface. (b) In the case of forward scattering the interaction is localized in momentum space. The gap is forming in the region where the energy of two quasi-particles is smaller than the interaction energy $2 \xi_\vk < g$.}
\label{pairFig}
\end{figure}

Here, we suggest that SC in vanishing doping levels is directly connected to quantum criticality. The connection is made explicit by employing a specific type of microscopic coupling between mobile electrons and structural modes of the lattice. We model for these modes, using the quantum Ising Hamiltonian, and obtain an effective electron-electron interaction that is qualitatively different from Eq. (\ref{inter}). Similar to the BCS case, the self-consistent equation for the gap function with this interaction can be solved analytically. The solutions represent a pairing mechanism that is significantly different from BCS, as illustrated in Fig. \ref{pairFig}. In our case $N(0)$ loses its pivotal role and the critical temperature $T_c$ can be independent of the chemical potential, so, in principal superconductivity can occur without a Fermi surface, i.e., in an insulator. Furthermore, the gap function can be a non monotonous function of the temperature when the mediating modes are in the critical regime.

The need for a non-adiabatic theory is emphasized by experimental results with strontium titanate, which was found to be superconducting at extremely low doping levels \cite{dilute1,dilute2}. Besides framing the problem as an observed phenomenon, SC in strontium titanate has attracted a lot of attention regarding interfaces with different materials \cite{interFeSe1,interFeSe1,interLAO1,interJason} and due to the polaronic behavior \cite{PolarEagle1,PolarEagle2,PolarDev, polar1}. Recent experiments in this material revealed an interesting effect of doping on the thermal conductivity \cite{therm} and peculiar pairing at the interface \cite{prepair}. A good understanding of the bulk SC, which has been pursued for many years \cite{sto1,sto2,sto3} and is still debated \cite{gor2016phonon,ruhman,gor2016back}, would be highly valuable. In Ref. \cite{JonaPrl}, FE modes close to a quantum critical point (QCP) were suggested as the source of SC and used to explain the vanishing SC for increased levels of doping. An unusual isotope effect was proposed as a method to study the phenomenology of the critical behavior \cite{prbIso} and was experimentally observed \cite{stucky}, supporting the connection of SC to the QCP. The coexistence of FE and SC was observed \cite{FEinDome} and strain was also proposed as a tuning parameter \cite{kirsty} and experimentally investigated \cite{pres}. More generally, a vast effort was focused on connections between SC and a QCP in the past decades. Nonetheless, new theoretical approaches \cite{Gqcp2} are presented and experimental studies \cite{Gqcp1} are performed every so often.

\emph{The model.}
The full Hamiltonian of the system is $H=H_m + H_e + H_{me}$, where $H_m$, $H_e$ and $H_{me}$ are the Hamiltonians for the FE modes, electrons and their interaction respectively. We start by deriving $H_m$ and then use its (mean field) solution, together with $H_{me}$, in order to get an effective electron electron interaction, which would be the basis for SC. 

In Ref. \cite{prbIso} it was already shown how FE modes can be described by the quantum Ising Hamiltonian. For completeness, we go over this derivation here. Consider a single unit cell, containing several ions where some are charged positively and other negatively. The configuration of the positions of the ions can have two degenerate energy minima while the most symmetric configuration happens to be a local maximum, i.e. a double-well potential. A quantization of the system reveals that the ground state of the system is an equal and symmetric superposition of the two minima and the first excited state is an antisymmetric superposition. Neglecting higher levels we can write the Hamiltonian for a single unit cell as $H={1 \over 2} \Gamma \sigma_x$, where $\sigma_x$ is the Pauli matrix, having the eigenstates $\ket{\uparrow_x}$ and $\ket{\downarrow_x}$ with eigenvalues $1$ and $-1$, respectively and $\Gamma$ is the excitation energy (or the tunneling frequency). 

The eigenstates of $\sigma_x$ represent symmetric and antisymmetric superpositions so the eigenstates of $\sigma_z$ imply the configuration of the ions is around one of the minima. These configurations entail an electric dipole, due to the different charge of the ions. The direction, and magnitude, of the dipole $\vec{d}$ is given by the details of the configuration on the minimum. The dipoles for the two states $\ket{\uparrow_z}  =(\ket{\uparrow_x} + \ket{\downarrow_x}) /\sqrt{2} $ and $\ket{\downarrow_z}=(\ket{\uparrow_x} - \ket{\downarrow_x}) /\sqrt{2} $ have the same magnitude and opposite directions, so we can write the dipole pertaining to these pair of minima as $\sigma^\alpha_z (i) \vec{d}$, where $i$ denotes the unit cell and $\alpha$ denotes the minima pair. In general, there would be other pairs of degenerate minima, with electric dipoles pointing in different directions, typically one pair for each spatial direction $\alpha = x,y,z$ (not to be confused with the index of the Pauli matrices $x,z$ which refer to the pseudo-spin direction).  


The electric field created by the dipole induces a coupling between different unit cells (and also a coupling to mobile electrons which we describe below). A general dipole-dipole interaction can be written as $H_{dd}=-{1 \over 2} \sum_{i,j,\alpha,\beta} J^{\alpha,\beta}_{i,j} \sigma^\alpha_z(i) \sigma^\beta_z(j)$, where $ J^{\alpha,\beta}_{i,j}$ is the interaction energy. For simplicity, let us assume $ J^{\alpha,\beta}_{i,j} \sim \delta^{\alpha,\beta}$, so different modes are decoupled. We consider now a single mode and suppress the mode index. Later, the effective interaction mediated by these modes will include a sum over them.
Together with the onsite energy we obtain the Hamiltonian for the FE modes in the form of the quantum Ising model
\begin{align} \label{Hs}
H_m=-{1 \over 2} \Gamma \sum_{i}\sigma_x(i) - {1 \over 2} \sum_{i,j} J_{i,j} \sigma_z(i) \sigma_z(j).
\end{align}
This model, which was investigated vastly \cite{suzuki2013,sachdevBook}, describes a quantum phase transition between a FE order $\expval{\sigma_z} \ne 0$ when $J \gg \Gamma $, and a paraelectric phase $\expval{\sigma_z} = 0$ when $J \ll \Gamma$, with $J$ being the scale of $J_{i,j}$. Using a mean field approximation one can obtain a solution for the Heisenberg operators $\sigma_z(\mathbf{q}) = \sum_{j} e^{i \mathbf{R_j} \cdot \mathbf{q}} \sigma_z(j )$ as $\sigma_z(\mathbf{q},t) = e^{ i \omega_\mathbf{q} t} \sigma_z(\mathbf{q},t=0)$ with  $\omega_\mathbf{q}= \sqrt{\Gamma \left(\Gamma - J_{\mathbf{q}} \right)}$, where $J_\mathbf{q} = \sum_j e^{i \mathbf{R_j} \cdot \mathbf{q}} J_{0,j}$ is the Fourier transform of the dipole-dipole interaction $J_{i,j}$ \cite{suzuki2013}. Since such an interaction is highly anisotropic and peaked at $q=0$, the dispersion relation has a minimum at $q=0$ and would depend mostly on $q_\alpha$, the longitudinal component of $\mathbf{q}$. At the QCP $\omega_0=\omega_{q=0}$ has to vanish and the critical regime of the system is defined as $\omega_0 < T$, with $T$ being the temperature.

In the standard treatment of electron-phonon coupling, namely the Fr\"olich Hamiltonian, the phonons are assumed to be harmonic structural modes. This assumption implies that the excitation levels are equidistant and thus can be described using bosonic creation and annihilation operators. Our model is valid in the opposite regime, where a strong anharmonicity, in the form of a double-well potential, allows one to neglect levels higher than the first excitation and leads to a pseudo-spin description. This description for the structural modes, formulated by the quantum Ising model, is suitable when the system is close to criticality \cite{sachdevBook}.

The interaction between electrons and the FE mode is given by $H_{me} = \sum_{i,j} c^\dagger_i c_i \sigma_z(j) \phi_{i,j}$, where $c^{(\dagger)}$ is the electronic annihilation (creation) operator and $\phi_{i,j}$ is the electric potential at site $i$ due to a dipole at site $j$. Similar to the dispersion $\omega_\vq$, this potential is strongly anisotropic. Moreover, its Fourier transform $\phi_\mathbf{q} = \sum_j e^{i \mathbf{R_j} \cdot \mathbf{q}} \phi_{0,j} \propto q_\alpha \abs{\mathbf{q}}^{-2}$, is peaked at $q=0$, in contrast to acoustic phonons, whose coupling to electrons vanishes there. 

The modes we are interested in are longitudinal, in the sense that the coupling is to the component of $\vq$ that is parallel to $\alpha$, which denotes the direction of the electric dipole. In the case of a broken continuous symmetry, there will be gapless Goldstone modes that are transverse. When the transition is at finite temperature and the ground state breaks a continuous symmetry, only these transverse modes remain soft. Here, we consider an Ising transition at zero temperature, so there are no Goldstone modes and at the QCP there is no gap. The longitudinal modes have much stronger dispersion but the difference in their frequency, compared to transverse modes, has to be at least $O(q)$ and typically it is $O(q^2)$ (see appendices \ref{model} and \ref{softLO} for details).

\emph{Deriving a self-consistent equation.}
Using $H_m$ and $H_{me}$, and treating the electronic density as an external source, we can write a solution for the FE Mastubara operators $ \sigma(\tau) = e^{ \tau H} \sigma e^{ - \tau H}$, as 
\begin{align} \label{sigma}
\sigma_z(\mathbf{q},\tau) &= \int_{-\beta}^\beta d^3k d\tau' D_\mathbf{q}(\tau-\tau') \Gamma c_\mathbf{k}^\dagger (\tau') c_\mathbf{k-q} (\tau') \phi_\mathbf{-q}
\end{align}
where $ c^{(\dagger)} (\tau) = e^{ \tau H} c^{(\dagger)} e^{ - \tau H}$ are the electronic Mastubara operators and 
$D_\mathbf{q}(\tau) =T \sum_\omega { - e^{i \omega \tau} \over \omega_\mathbf{q}^2 + \omega^2} $
is the Matsubara Green's function for a single FE mode (see appendix \ref{FEprop} for details). 

Once the solution for the FE modes is given, a standard procedure can be followed to introduce an effective electron-electron interaction and obtain a self-consistent equation for the electronic gap function \cite{mahan}. The main steps are sketched below, using the method of solving the equations of motions for the electronic Green's functions (For a more detailed calculation see appendix \ref{selfCons}). Assuming a simple one band model for the electrons, $H_e= \int d\mathbf{k} \xi_\mathbf{k} c_\mathbf{k}^\dagger c_\mathbf{k} $, where $\xi_\mathbf{k}$ is the energy, the time derivative of the electronic Matsubara operator is given by
\begin{align} \label{C}
-\partial_\tau c_\mathbf{k}(\tau) &= - [H, c_\mathbf{k}(\tau)] = \xi_\mathbf{k} c_\mathbf{k} + \int d\mathbf{q} \phi_\mathbf{-q} c_\mathbf{k+q} \sigma(\mathbf{q}) \nonumber\\
&=\xi_\mathbf{k} c_\mathbf{k} + \int d\mathbf{p} d\mathbf{q} d\tau' V(\mathbf{q},\tau-\tau') c_\mathbf{k+q} c_\mathbf{p}^\dagger (\tau') c_\mathbf{p-q} (\tau'), 
\end{align}
where in the second line we have inserted Eq. (\ref{sigma}) and
$V(q,\tau) = \sum_\alpha D^\alpha_\mathbf{q}(\tau) \Gamma \abs{\phi_\mathbf{q}^\alpha}^2 $ is an effective retarded interaction with a summation over FE modes.

We define the Matsubara Green's functions in momentum space $G(\mathbf{k} ,\tau) = T_\tau \left\langle c_\mathbf{k}(\tau) c^{\dagger}_\mathbf{k}(0) \right\rangle$, $F^{\dagger}(\mathbf{k} ,\tau) = T_\tau \left\langle c^{\dagger}_\mathbf{k}(\tau) c^{\dagger}_\mathbf{-k}(0) \right\rangle$, where $T_\tau$ is the Matsubara time ordering operator. Their time derivatives, $\partial_\tau G(\mathbf{k} ,\tau)$ and $\partial_\tau F^{\dagger}(\mathbf{k} ,\tau)$, after one inserts Eq. (\ref{C}), include 4-point functions $\expval{c c^{\dagger} c c^{\dagger}}$ and $\expval{c^{\dagger} c^{\dagger} c c^{\dagger}}$, which can be approximated by introducing a mean field $\Delta \propto \expval{c c }$. A Fourier transform to Matsubara frequency $G(\tau) =T \sum_{\omega} e^{-i \omega \tau} G(\omega),F^{\dagger}(\tau) =T \sum_{\omega} e^{-i \omega \tau} F^{\dagger}(\omega)$, with $\omega = \pi T (2 n + 1) $, and a definition of a gap function
\begin{align} \label{gap}
\Delta(\mathbf{k} ,\omega) &=T \sum_{\mathbf{q} ,\omega'} 
V(\mathbf{q},\omega-\omega') F(\mathbf{k+q} ,\omega'), 
\end{align}
where
\begin{align} \label{pot}
V(\vq,\omega) = \int_{-\beta}^\beta d\tau V(q,\tau) e^{i \omega \tau} =- \sum_\alpha { \Gamma \abs{\phi_\mathbf{q}^\alpha}^2 \over \left( \omega^\alpha_\vq \right)^2 + \omega^2}, 
\end{align}
results in two coupled equations for $ G(\mathbf{k} , \omega)$ and $ F^{\dagger}(\mathbf{k} , \omega )$, which can be solved analytically. Inserting a solution for $F^{\dagger}(\mathbf{k} ,\omega)$, in terms of $\Delta(\mathbf{k} ,\omega) $, back into Eq. (\ref{gap}), yields a self consistent equation 
\begin{align} \label{selfConst}
\Delta(\mathbf{k} ,\omega) = -T \sum_{\mathbf{k'},\omega'} 
V(\mathbf{k-k'},\omega-\omega') { \Delta(\mathbf{k'} ,\omega') \over \Delta^2(\mathbf{k'} ,\omega') + \omega'^2 + \xi^2_\mathbf{k'} }. 
\end{align}
In the usual Eliashberg treatment \cite{eliashberg}, which relies on Migdal’s theorem \cite{migdal} for neglecting terms of order $O(\omega_D / E_f)$, the frequency dependence of the gap function comes from the self-energy of the electrons. In contrast, here, the energy scales of the FE modes are comparable to those of the mobile electrons and it is their dynamics that can lead to a significant frequency dependence. We will use the physical properties of the FE modes in order to approximate $V(\mathbf{q},\omega)$ such that Eq. (\ref{selfConst}) can be solved. 

\emph{Solving the self-consistent equation.}
We start with the frequency dependence. To this end it is convenient to write the Matsubara frequencies as $\omega_n = T \tilde{\omega}_n$ with $\tilde{\omega}_n = \pi (2 n + 1) $ and to note that since the interaction in Eq. (\ref{selfConst}) is a function of the difference between two fermionic frequencies, it includes a term $V(\mathbf{q},\omega=0) \propto \omega_\vq^{-2}$ while the rest of the terms are $\propto T^{-2}$. At the critical regime $\omega_{q=0} < T$, so if there is a sufficient range of $\vq$ where $\omega_{\vq} \ll T$, the term $V(\mathbf{q}) = V(\mathbf{q},\omega=0) $ will dominate the sum. Neglecting terms with $n \ne 0$, we have $V(\mathbf{q},\omega_n) =V(\mathbf{q}) \delta_n$, where $\delta_n$ is a Kronecker delta. Thus the frequency sum in Eq.  (\ref{selfConst}) is trivial and we get
\begin{align} \label{selfConst2}
\Delta_n(\mathbf{k} ) = \sum_{\mathbf{k'}} V(\mathbf{k-k'})
{  -T \Delta_n(\mathbf{k'} ) \over \Delta_n^2(\mathbf{k'}) + T^2  \tilde{\omega}_n^2 + \xi^2_\mathbf{k'} },
\end{align}
with $\Delta_n(\mathbf{k} )= \Delta(\mathbf{k},\omega_n )$. We now have a separate equation for each frequency component $\Delta_n(\mathbf{k} )$. In the opposite regime, $\omega_{\vq} \gg T$, one can neglect the frequency dependency of $V(\mathbf{q},\omega)$ which implies $\Delta(\mathbf{k} ,\omega)= \Delta(\mathbf{k})$ is also frequency independent. Then, the frequency sum in Eq. (\ref{selfConst}) can be done and the typical form of the equation is obtained, $ \Delta(\mathbf{k}) =\sum_{\mathbf{k'}} V(\mathbf{k-k'}) \Delta(\mathbf{k'}) \tanh\left( \frac{E_{\mathbf{k'}}}{2T}\right) / 2 E_{\mathbf{k'}}$, with $E_\mathbf{k} = \sqrt{ \Delta^2(\mathbf{k}) + \xi^2_\mathbf{k}}$.

We now turn to the momentum dependence of the interaction. Note that while $\phi_\mathbf{q}$, appearing in the numerator of Eq. (\ref{pot}) is peaked at $q=0$, $\omega_\mathbf{q}$ in the denominator has its minimum there. So it is plausible to think that $V(\mathbf{q})$ would be strongly peaked at $q=0$. If the width of this peak, which depends on the properties of the FE modes, is small compared to the other momentum dependency, due to $\xi_\mathbf{k}$, then it can be approximated by a Dirac delta $V(\mathbf{q}) \simeq - g \delta(q)$, where $g>0$ is a coupling constant. This limit, where forward scattering is the main process for electron pairing, was discussed in a wide range of systems such as the cuprates \cite{ForwScatCupYang, ForwScatCupSD,ForwScatCupMD,ForwScatCup1}, FeSe interface \cite{ForwScatFeSe1,ForwScatFeSe2} and iron pnictides \cite{ForwScatIron}. It was used to explain anisotropies in the gap function, leading to different symmetries \cite{ForwScatCupSD,ForwScatCupMD, ForwScatIron}, enhancement of $T_c$ \cite{ForwScatFeSe2}, pseudogap behavior \cite{ForwScatCupYang} and a broadening of the phonon line shape \cite{ForwScatFeSe2}. It results in momentum decoupling  \cite{ForwScatCupMD} which makes the momentum sum in Eq. (\ref{selfConst2}) trivial and we obtain a separate equation for each component with the solutions
\begin{align} \label{solution1}
\Delta_n(\mathbf{k} ) = \Re \sqrt{g T - T^2 \tilde{\omega}_n^2 - \xi^2_\mathbf{k} },
\end{align} 
$T^\pm_c(n,\mathbf{k}) = {g \pm \sqrt{g^2 - 4 \tilde{\omega}_n^2 \xi^2_\mathbf{k}} \over 2 \tilde{\omega}_n^2 }.$
For $\xi_\mathbf{k} = 0$, we have $\Delta_n(\mathbf{k} ) \ne 0$ for $T < T_c(n,\mathbf{k}_F) = g/ \tilde{\omega}_n^2$, similar to the usual understanding of a critical temperature. Away from the Fermi surface we have two critical temperatures and $\Delta_n(\mathbf{k} ) \ne 0$ in the range $T^-_c < T < T^+_c$. As long as the normal state is metallic, $T^-_c$ might be irrelevant, since other components of the gap can be finite below it. For an insulator, this temperature might indicate an insulator-superconducting transition, driven by thermally excited carriers.
  
\begin{figure}
\centering
\includegraphics[width=0.234\textwidth]{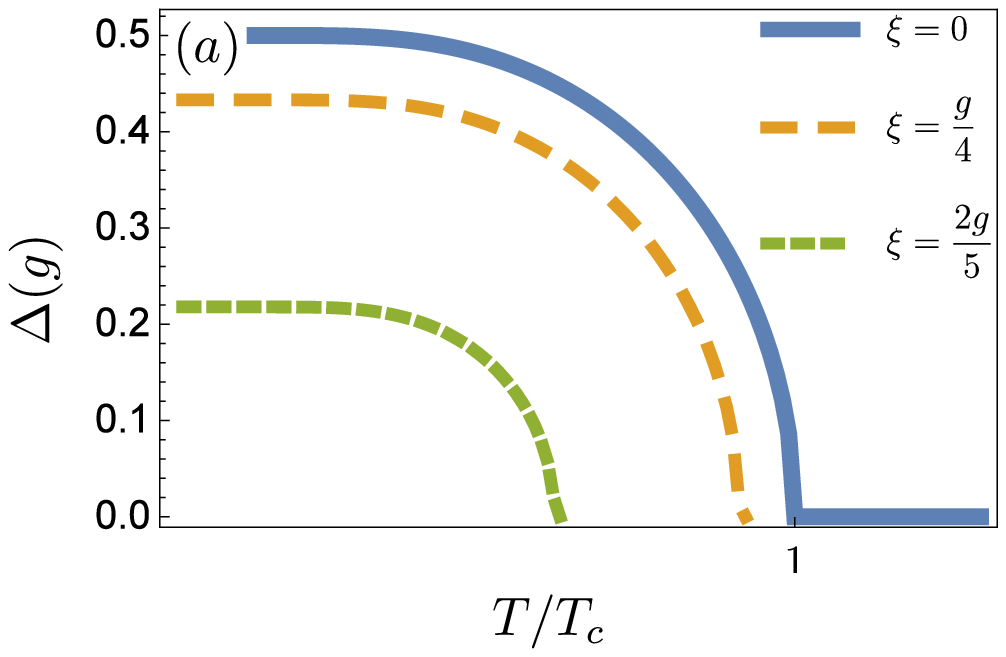}
\includegraphics[width=0.234\textwidth]{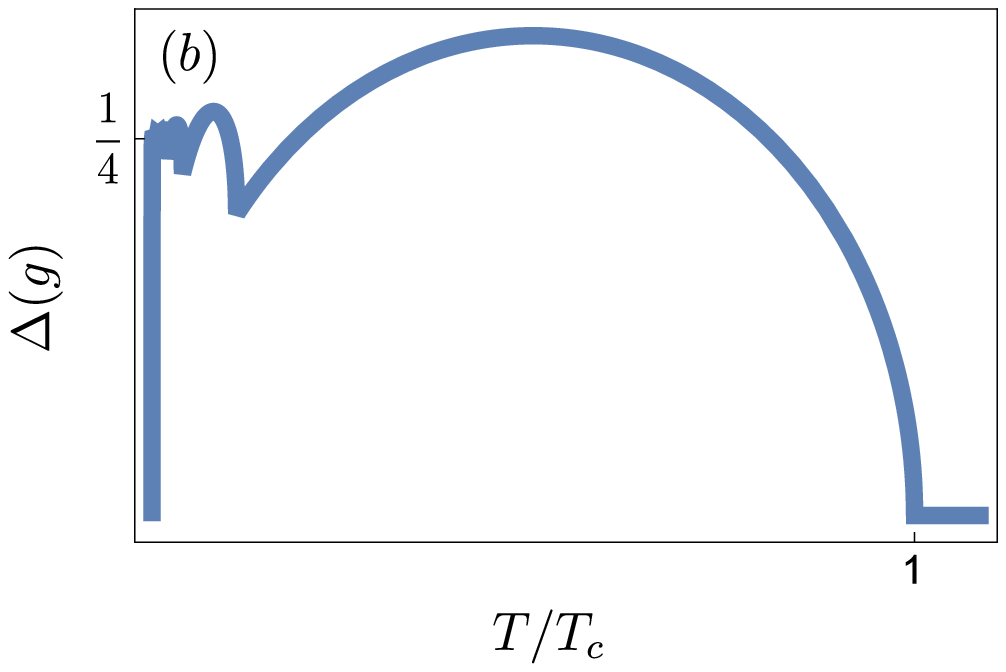}

\includegraphics[width=0.234\textwidth]{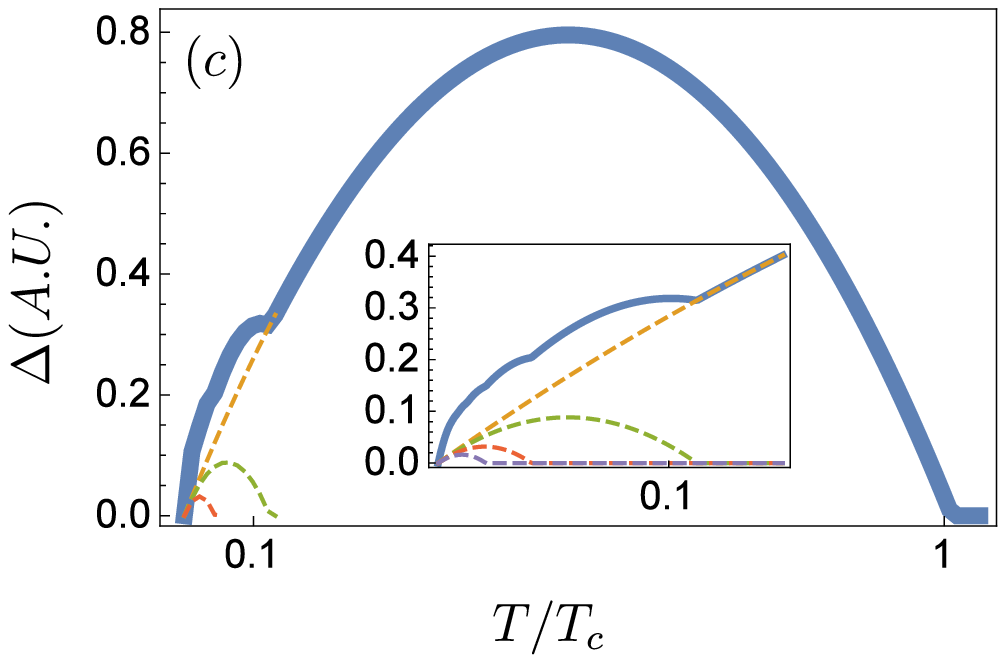}
\includegraphics[width=0.234\textwidth]{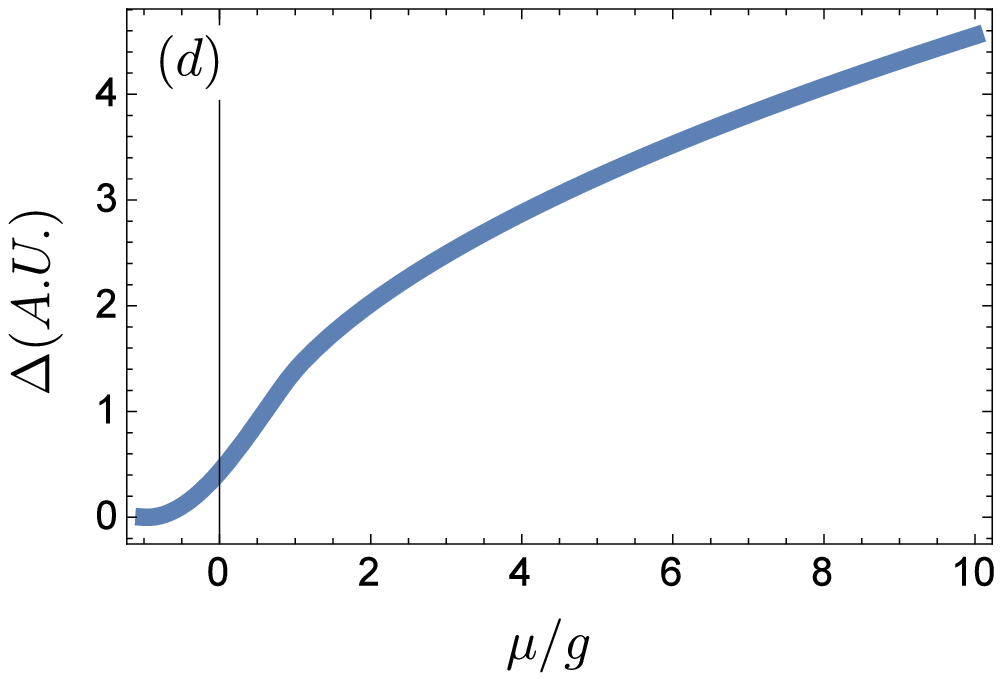}

\caption{The dependency of the gap function $\Delta$ on temperature and doping. (a) For $\omega_{\vq} \gg T$, a numerical solution of the gap equation is shown for different values of $\xi_\mathbf{k}$. (b) For $\omega_{\vq} \ll T$, the sum over the analytical solutions in Eq. (\ref{solution1}) at the Fermi surface, where each one vanishes as $\sqrt{T}$ for low temperatures, shows a finite result $\lim_{T \to 0} \sum_n \Delta_n(\mathbf{k}_F ) = g/4$, due to more terms activated at $T_c(n) = g/ \tilde{\omega}_n^2$. (c) Integration over $\xi_\mathbf{k}$ using a 3D density of states, for $\omega_{\vq} \ll T$, is showing the behavior described by Eq. (\ref{del3}). High frequency components are shown as dashed lines (more clearly in the inset). (d) The dependency of $\Delta$ on doping showing a finite value for $\mu < 0$, i.e. insulator. (only the case of $\omega_{\vq} \gg T$ is plotted since the plot for $\omega_{\vq} \ll T$ is similar). In panels (c) and (d) the integration over $\xi_\mathbf{k}$ requires introducing another parameter related to the density of state.}
\label{gapFig}
\end{figure}

In the case $\omega_{\vq} \gg T$, using $V(\mathbf{q}) \simeq -g \delta(q)$ yields a separate (transcendental) equation $ 2 E_{\mathbf{k}} = g  \tanh\left( \frac{E_{\mathbf{k}}}{2T}\right)  $ for each $E_\mathbf{k} $. At the Fermi surface the critical temperature is given by $T_c(\mathbf{k}_F) = g / 4$, and at $T=0$ the gap function is \cite{ForwScatCupYang}
 \begin{align} \label{solution2}
\Delta(\mathbf{k} ) = \sqrt{g^2/4 - \xi^2_\mathbf{k} }.
\end{align}
A general solution for $\Delta(\mathbf{k} )$, as a function of $T$ and $\xi_\mathbf{k}$ is shown in Fig. \ref{gapFig} (a).

As a comparison to these results, one can consider the opposite regime, of interaction that is momentum independent $V(\mathbf{q}) \simeq -g $. In the case $\omega_{\vq} \gg T$, the momentum integral would diverge and the standard procedure, introducing a cutoff $\omega_D$, results in $T_c = \omega_D e^{- 1/ g N(0)}$. In the case of Eq. (\ref{selfConst2}) the momentum integral does not diverge and can be done analytically, using a typical density of states. The resulting gap function is non zero only for large temperature. A physical interpretation of this scenario is an interesting question, which we do not address in this work.

\emph{Discussion.} The state described by Eq. (\ref{solution2}) has some similarities to the BCS case. The density of states $N(\xi) \propto \Theta(g- 2 \abs{\xi}) (\partial \xi_\mathbf{k}  / \partial \mathbf{k}|_{\xi_\mathbf{k}  =\xi})^{-1}$ has a gap of $E_g = g = 4 T_c$ but no square root singularity. The ratio $E_g/T_c$ differs by a factor of 1.14 from the BCS result, coming from the lack of integration over momentum/energy. The case of Eq. (\ref{solution1}) is rather different. It is valid only in the critical regime of the FE modes, which is mostly $T>0$. It does include a single point in the parameter space with $T=0$, namely the QCP, but at this point $\omega_0 \rightarrow 0$ and thus $g \rightarrow \infty$. So it is not straight forward to take the limit $T\rightarrow 0$ in order to obtain, for example, the retarded Green's function, spectral function etc. This might imply that Eq. (\ref{solution1}) does not describe any ground state.

The results in Eqs. (\ref{solution1})  and (\ref{solution2}) do not depend directly on the density of carriers, in strong contrast to the corresponding expression in BCS theory where $N(0)$ appears in the exponent. This can be attributed to the infinite range of the forward scattering process. The solutions in Eqs. (\ref{solution1}) and (\ref{solution2}) show possible pairing channels but any observed phenomena, such as persistent current, Meissner effect, Josephson effect etc., might depend on how many channels contribute. In order to study the possible dependence on the doping level and temperature we consider the quantity $\Delta = \sum_{n, \mathbf{k}} \Delta_n(\mathbf{k} )$, which is plotted in Figs. \ref{gapFig} (c) and (d) as a function of temperature and chemical potential $\mu$ \footnote{In order to avoid considering the non trivial relation between the Fermi energy and carrier density at finite temperature, we use the grand canonical ensemble, which allows us also to consider negative $\mu$.}, respectively. The results can be observed experimentally by measuring, for example, the critical current. The qualitative behavior can be inferred at some limits. For a high level of doping $\mu \gg g$ \footnote{We compare $\mu$ to $g$, so the regime of a high doping level here can correspond to a low level in other contexts and in any case should not be associated with the vanishing SC at increased doping (the other side of the SC dome)} we have $\Delta \propto N(0)  ( T_c^2 - T_2)$ in the case of  Eq. (\ref{solution2}) and 
 \begin{align} \label{del3}
\Delta \propto N(0)  T T_c  \left( \sqrt{T_c \over T} -  \sqrt{T \over T_c}\right)
\end{align}
in the case of  Eq. (\ref{solution1}), where in both cases $T_c \propto g$. For a low level of doping $\mu \ll g$ we have, in 3 dimensions, $\Delta \propto  E_F/T_c + 1$, which implies SC does not vanish for $\mu=0$ and even for $\mu<0$, i.e. an insulator. This result is quite generic, since for $g > - \mu > 0$ it can be energetically favored to excite a pair of carriers and allow them to form a bound state. However, this might require extremely strong coupling and low temperature.

The pairing mechanism introduced in this work has the double benefit of being derived from a microscopic model and resulting in unique features that can be experimentally observed. While relying on a physical picture that is different from BCS, this theory still possesses a major advantage of BCS, namely it is analytically tractable. The concrete connection between QCP and SC can be employed as a tool for a theoretical analysis in a wide range of systems. The results might be used to explain pre-pairing observations or the pseudo-gap phenomenon. 

{\em Acknowledgments} I would like to thank J. M. Edge, T. Kvorning, T. H. Hansson, S. Pershoguba and A.V. Balatsky for helpful discussions. Parts of the results presented in this paper were obtained in collaboration with C. Triola and E. Langmann.

\begin{widetext}

\appendix
\section*{Appendices}
\renewcommand{\thesubsection}{\Alph{subsection}}
 


\subsection{The model} \label{model}

Our system comprised of itinerant electrons and electric dipoles that are the source of possible ferroelectricity. The Hamiltonian of the combined system is $H = H_e + H_m + H_{me} + H_{ee}$ where $H_e$, $H_m$,  $H_{me}$ and $H_{ee}$ are the Hamiltonians for the electrons, the dipoles, their interaction and the direct electron-electron Coulomb interaction. We include here $H_{ee}$ for completeness, even tough it is not included in the main text. The coulomb repulsion lessen superconductivity but has no qualitative effect and including in make the expressions more cumbersome. 

The Hamiltonian for a single mode and its coupling to the electrons were derived in the main text. Here we keep the sum over modes throughout the derivation. As mentioned in the main text the modes are assumed to have no interaction between the different directions. In each unit cell $i$, reside three pseudo-spins $\vec{\sigma}^{\alpha=x,y,z}(i)$ representing the polarization, i.e. an electric dipole, in the $x$, $y$ and $z$ directions. Do not confuse the index $\alpha$, which refer to the spatial direction, with the  pseudo-spin index referring to different Pauli matrices. The assumption that different modes do not interact means the polarization in one direction, say $x$, does not depend on the polarization in another, say $y$, so we do not have term such as $ \sigma^x_z \sigma^y_z$.

The real space Hamiltonians are given by
\begin{align} \label{H}
H_m &= -{1 \over 2}\Gamma\sum_{i,\alpha} \sigma^\alpha_x(i) - {1 \over 4} \sum_{i,j,\alpha} J^\alpha_{i,j}\sigma^\alpha_z (i) \sigma^\alpha_z (j), \\
H_e &= - \sum_{i,j,s} t_{i,j} c^{\dagger,s}_i c^s_j, \\
H_{ee} &= \sum_{i,j,s,s'} V^c_{i,j} c^{\dagger,s}_i c^s_i c^{\dagger,s'}_j c^s_j,\\
H_{me} &= \sum_{i,j,s,\alpha} c^{\dagger,s}_i c^s_i \sigma^\alpha_z(j) \phi^\alpha_{i,j}.
\end{align}
Here, $\Gamma$, $J^\alpha_{i,j}$, $t_{i,j}$, $V^c_{i,j}$ and $\phi_{i,j}$ are the onsite tunneling frequency for the dipole, dipole-dipole interaction, hopping amplitude for the electrons, Coulomb interaction and the electron-dipole interaction respectively. $i$ and $j$ are site indices and $s$ is the spin. Most of the parameters of the Hamiltonians are assumed to have a standard form and we do not specify an explicit expression. The only non common parameter, $\phi^\alpha_{i,j}$ is given by the potential induced on an electron at site $i$ by an electric dipole in the direction $\alpha$  located at site $j$
\begin{equation} 
\phi^\alpha_{i,j} \propto   {   (\vR_i -\vR_j)_\alpha \over |\vR_i - \vR_j|^3},
\end{equation}
where $\vR_i$ is the position of site $i$ and $(\vR_i -\vR_j)_\alpha $ is the $\alpha$ component of that position.
 
Let us transform everything to momentum space by using $\sigma^\alpha_z(j) = \int d^3q  e^{i \mathbf{R_j} \cdot \mathbf{q}} \sigma^\alpha_z( \mathbf{q})$ and $ c_j =  \int d^3k  e^{i \mathbf{R_j} \cdot \mathbf{k}}   c_\mathbf{k}$:
\begin{align} \label{Hk}
H_m &= -{1 \over 2} \sum_{\alpha}\Gamma \sigma^\alpha_x(\mathbf{q} =0) - {1 \over 4} \sum_{\mathbf{q},\alpha} J^\alpha_{\mathbf{q}} \sigma^\alpha_z (\mathbf{q}) \sigma^\alpha_z (- \mathbf{q}) \\
H_e &=  \sum_{\mathbf{k},s}  \xi_\mathbf{k} c^{\dagger,s}_\mathbf{k} c^s_\mathbf{k} \\
H_{ee} &=  \sum_{\vk,\vq,\vp,s,s'} V^c_{\vq} c^{\dagger,s}_\vk c^s_{\vk+\vq} c^{\dagger,s'}_\vp c^{s'}_{\vp-\vq}\\
H_{me} &= \sum_{\mathbf{k},\mathbf{q},s,\alpha}  c^{\dagger,s}_\mathbf{k} c^s_\mathbf{k+q} \sigma^\alpha_z(\mathbf{q}) \phi^\alpha_\mathbf{q} =  \sum_{\mathbf{q}}  \hat{n}(\mathbf{q}) \sigma^\alpha_z(\mathbf{q}) \phi^\alpha_\mathbf{q} 
\end{align}
where $J^\alpha_{\mathbf{q}}$,  $\xi_\mathbf{k}$, $V^c_{q}$ and $\phi^\alpha_\mathbf{q}$ are the Fourier transforms of $ J^\alpha_{i,j}$, $t_{i,j}$, $V^c_{i,j}$ and $  \phi^\alpha_{i,j}$ respectively. The dipole-electron interaction is now given by
\begin{equation} \label{dip}
\phi^\alpha_\vq \propto   {   q_\alpha  \over q^2}.
\end{equation}
This equation implies that only longitudinal modes couple to the electrons and not the transverse ones, in agreement with most treatments of coupling optical modes to electrons. The meaning of longitudinal and transverse here is slightly different, even though the definitions are essentially the same. In a typical discussion of this issue, one consider some wave vector $\vq$ and then study ions displacements, or polarization vector, that are parallel or perpendicular to $\vq$. These modes are usually considered to have different properties and thus are treated separately. Here, we fix the direction of polarization from the start, using the index $\alpha=x,y,z$, and refer to the component of a wave vector $\vq$ that is parallel (perpendicular) to $\alpha$ as longitudinal (transverse).

\subsection{The propagator of the FE modes} \label{FEprop}
In order to get the equations of motion for the dipoles it can be easier to work in momentum space. To this end let us derive the commutation relation of the momentum space Pauli matrices $\sigma^\alpha_a(\mathbf{k} )$, where $a=x,y,z$ is the index denoting which Pauli matrix, and $\alpha$ denotes the direction of polarization, using the usual commutation relation in real space $ \left[ \sigma^\alpha_a(j ), \sigma^\beta_b(j )  \right] = 2 i \varepsilon_{a b c}\sigma^\alpha_c(j ) \delta_{i,j} \delta_{\alpha,\beta}$. We have  
\begin{align}
\left[ \sigma^\alpha_a(\mathbf{k} ), \sigma^\beta_b(\mathbf{q} ) \right] &= N^{-1}  \sum_{i,j}  e^{i \mathbf{R_j}  \cdot \mathbf{k} + i \mathbf{R_i}  \cdot \mathbf{q} } \left[ \sigma^\alpha_a(j ), \sigma^\beta_b(j )  \right]  \nonumber \\   
&=  N^{-1}  \sum_{i,j}  e^{i \mathbf{R_j}  \cdot \mathbf{k} + i \mathbf{R_i}  \cdot \mathbf{q} } 2 i \varepsilon_{a b c}\sigma^\alpha_c(j ) \delta_{i,j}  \delta_{\alpha,\beta} \nonumber \\   
&=  N^{-1}  \sum_{j}  e^{i \mathbf{R_j}  \cdot (\mathbf{k} + \mathbf{q} } 2 i \varepsilon_{a b c}\sigma^\alpha_c(j ) \delta_{\alpha,\beta} = 2 i \varepsilon_{a b c}\sigma^\alpha_c(\mathbf{k} + \mathbf{q}  ) \delta_{\alpha,\beta}
\end{align}

Now we write the operators as Matsubara operators 
\begin{equation} 
 \sigma(\tau) = e^{ \tau H}  \sigma e^{ - \tau H}.
\end{equation}
so that in general $\partial_\tau \sigma(\tau) =   [H, \sigma]$ and in our case we have :
\begin{equation} 
\partial_\tau \sigma^\alpha_z(\mathbf{k},\tau) =  [H, \sigma^\alpha_z(\mathbf{k},\tau)]   =   i \Gamma \sigma^\alpha_y(\mathbf{k},\tau).
\end{equation}
The second order will give us
\begin{align} \label{d2z}
\partial_\tau^2 \sigma^\alpha_z(\mathbf{k},\tau) &=    i \Gamma [H, \sigma^\alpha_y(\mathbf{k},\tau)] \nonumber \\   
&=   \Gamma \left(  \Gamma \sigma^\alpha_z(\mathbf{k},\tau)   - {1 \over 2} \sum_{\mathbf{q}} J^\alpha_{\mathbf{q}} \left(
\sigma^\alpha_x (\mathbf{k+q},\tau) \sigma^\alpha_z (- \mathbf{q},\tau) +
\sigma^\alpha_z (\mathbf{q},\tau) \sigma^\alpha_x ( \mathbf{k-q},\tau)
\right)\right. \nonumber \\   
 &\left.
+\sum_{\mathbf{q}} \hat{n}(\mathbf{q},\tau) \sigma^\alpha_x(\mathbf{q+k},\tau) \phi^\alpha_\mathbf{q} 
\right) \end{align}

In order to solve this equation we assume mean field for $\sigma^\alpha_x$  so $\sigma^\alpha_x(\mathbf{k}) = \delta(\mathbf{k}) $. This is valid when $ \Gamma \gg J$ so the system is deep in the paramagnetic region. We get
\begin{align} \label{d2zB}
\partial_\tau^2 \sigma^\alpha_z(\mathbf{k},\tau) &=  \Gamma \left(\Gamma - J^\alpha_{\mathbf{k}} \right)  \sigma^\alpha_z ( \mathbf{k},\tau) 
+ \Gamma \hat{n}(\mathbf{-k})  \phi^\alpha_\mathbf{-k}
\end{align}
This equation can be solved formally as 
\begin{align}  \label{sigz}
 \sigma^\alpha_z(\mathbf{k},\tau) &=  \int_{-\beta}^\beta d\tau' D^\alpha_\mathbf{k}(\tau-\tau')  \Gamma \hat{n}(\mathbf{-k},\tau')    \phi^\alpha_\mathbf{-k}
\end{align}
where $D^\alpha_\mathbf{k}(\tau)$ satisfies 
\begin{align} 
\left(\partial_\tau^2 -\left( \omega^\alpha_\mathbf{k}\right)^2\right) D_\mathbf{k}^\alpha(\tau) =  \delta(\tau)
\end{align}
with $\omega^\alpha_\mathbf{k} = \sqrt{\Gamma \left(\Gamma - J^\alpha_{\mathbf{k}} \right)}$. We can write a solution as 
\begin{align} 
 D^\alpha_\mathbf{k}(\tau) = T \sum_\omega { - e^{i \omega \tau} \over \left( \omega^\alpha_\mathbf{k}\right)^2 + \omega^2} .
\end{align}

\subsection{Deriving a self consistent equation for the gap function} \label{selfCons}

The equations for the electron Matsubara operators are 
\begin{align} 
\partial_\tau c^{s}_\mathbf{k} &=   [H, c^{s}_\mathbf{k}]   = -  \xi_\mathbf{k}  c^{s}_\mathbf{k}
-\sum_{\mathbf{q,p},s'} V^c_\mathbf{q} c^{\dagger,s'}_\mathbf{p} c^{s'}_\mathbf{p-q}  c^{s}_\mathbf{k+q}
 -  \sum_{\mathbf{q},\alpha}  c^{s}_\mathbf{k+q} \sigma^\alpha_z(\mathbf{q}) \phi^\alpha_\mathbf{q}, \\
\partial_\tau c^{\dagger,s}_\mathbf{k} &=  [H, c^{\dagger,s}_\mathbf{k}]   =  \xi_\mathbf{k}  c^{\dagger,s}_\mathbf{k} +
\sum_{\mathbf{q,p},s'} V^c_\mathbf{q} c^{\dagger,s'}_\mathbf{p} c^{s'}_\mathbf{p-q} c^{\dagger,s}_\mathbf{k-q}
 + \sum_{\mathbf{q},\alpha}  c^{\dagger,s}_\mathbf{k-q} \sigma^\alpha_z(\mathbf{q}) \phi^\alpha_\mathbf{q}.
\end{align} 
Inserting the expression for $\sigma^\alpha_z(\mathbf{q},\tau) $ from Eq. (\ref{sigz}) we have
\begin{align} 
-\partial_\tau c^{s}_\mathbf{k}(\tau) &=  \xi_\mathbf{k}  c^{s}_\mathbf{k}(\tau) +
\sum_{\mathbf{q,p},s'} V^c_\mathbf{q} c^{\dagger,s'}_\mathbf{p}(\tau) c^{s'}_\mathbf{p-q}(\tau) c^{s}_\mathbf{k+q}(\tau)\nonumber \\
&+ \sum_{\mathbf{q,p},s',\alpha}  
  \int_{-\beta}^\beta d\tau' D^\alpha_\mathbf{q}(\tau-\tau')  \Gamma  \phi^\alpha_\mathbf{-q}  \phi^\alpha_\mathbf{q}
  c^{s}_\mathbf{k+q}(\tau)  c^{\dagger,s'}_\mathbf{p}(\tau') c^{s'}_\mathbf{p-q} (\tau') \nonumber \\
 &=  \xi_\mathbf{k}  c^{s}_\mathbf{k}(t) +
 \sum_{\mathbf{q,p},s'}  
  \int_{-\beta}^\beta d\tau' V(q,\tau-\tau')
  c^{s}_\mathbf{k+q}(\tau)  c^{\dagger,s'}_\mathbf{p}(\tau') c^{s'}_\mathbf{p-q} (\tau')  , \\
\partial_\tau c^{\dagger,s}_\mathbf{k}(\tau)
 &=  \xi_\mathbf{k}  c^{\dagger,s}_\mathbf{k}(\tau) +
\sum_{\mathbf{q,p},s'} V^c_\mathbf{q} c^{\dagger,s'}_\mathbf{p}(\tau) c^{s'}_\mathbf{p-q}(\tau) c^{\dagger,s}_\mathbf{k-q}(\tau)
            \nonumber \\
 & +  \sum_{\mathbf{q,k'},s',\alpha}    \int_{-\beta}^\beta d\tau' D^\alpha_\mathbf{q}(\tau-\tau')   \Gamma  \phi^\alpha_\mathbf{-q} \phi^\alpha_\mathbf{q}
  c^{\dagger,s}_\mathbf{k-q}(\tau)  c^{\dagger,s'}_\mathbf{k'}(\tau') c^{s'}_\mathbf{k'-q} (\tau')\nonumber \\
 &=  \xi_\mathbf{k}  c^{\dagger,s}_\mathbf{k}(\tau) +  \sum_{\mathbf{q,k'},s'}  
  \int_{-\beta}^\beta d\tau' V(q,\tau-\tau')c^{\dagger,s}_\mathbf{k-q}(\tau)  c^{\dagger,s'}_\mathbf{k'}(\tau') c^{s'}_\mathbf{k'-q} (\tau').
\end{align} 
where
\begin{align} 
 V(q,\tau) = \sum_{\alpha}D^\alpha_\mathbf{q}(\tau)    \Gamma    \abs{ \phi^\alpha_\mathbf{q}}^2 + V^c_{p} \delta(\tau) 
\end{align} 

Let us define the momentum space green functions:
\begin{align} \label{G}
G^{s,s'}(\mathbf{k} ,\tau) &=   T_\tau  \left\langle c^{s}_\mathbf{k}(\tau) c^{\dagger,s'}_\mathbf{k}(0) \right\rangle,\\ \label{Fdag}
F^{\dagger,s,s'}(\mathbf{k} ,\tau) &=   T_\tau  \left\langle c^{\dagger,s}_\mathbf{k}(\tau) c^{\dagger,s'}_\mathbf{-k}(0) \right\rangle,
\\ \label{F}
F^{s,s'}(\mathbf{k} ,\tau) &=   T_\tau  \left\langle c^{s}_\mathbf{k}(\tau) c^{s'}_\mathbf{-k}(0) \right\rangle.
\end{align}
The derivation with respect to $\tau$ yield 
\begin{align} 
\partial_\tau G^{s,s'}(\mathbf{k} ,\tau) &=\delta(\tau)  \delta_{s,s'}+   T_\tau  \left\langle \partial_\tau c^{s}_\mathbf{k}(\tau) c^{\dagger,s'}_\mathbf{k}(0) \right\rangle
 ,\\
\partial_\tau F^{\dagger,s,s'}(\mathbf{k} ,\tau) &=   T_\tau  \left\langle \partial_\tau c^{\dagger,s}_\mathbf{k}(\tau) c^{\dagger,s'}_\mathbf{-k}(0) \right\rangle
\end{align}

Now we can insert $\partial_\tau c^{s}_\mathbf{k}(t) $ and $\partial_t c^{\dagger,s}_\mathbf{k}(\tau)$ to get
\begin{align} 
\partial_\tau G^{s,s'}(\mathbf{k} ,\tau) &= 
\delta(\tau)  \delta_{s,s'} -  \xi_\mathbf{k}  G^{s,s'}(\mathbf{k} ,\tau) \nonumber \\
 &- \sum_{\mathbf{q,k'},r}  
  \int_{-\beta}^\beta d\tau' V(q,\tau-\tau') 
   T_\tau  \left\langle c^{s}_\mathbf{k+q}(\tau)  c^{\dagger,r}_\mathbf{k'}(\tau') c^r_\mathbf{k'-q} (\tau')  c^{\dagger,s'}_\mathbf{k}(0) \right\rangle , \label{G1}\\
\partial_\tau F^{\dagger,s,s'}(\mathbf{k} ,\tau) &= 
  \xi_\mathbf{k}  F^{\dagger,s,s'}(\mathbf{k} ,\tau) \nonumber \\
 & +   \sum_{\mathbf{q,k'},r}  
   \int_{-\beta}^\beta d\tau' V(q,\tau-\tau') 
  T_\tau  \left\langle  c^{\dagger,s}_\mathbf{k-q}(\tau)  c^{\dagger,r}_\mathbf{k'}(\tau') c^r_\mathbf{k'-q} (\tau') c^{\dagger,s'}_\mathbf{-k}(0) \right\rangle \label{F1}
\end{align}
The expressions above contain two 4-point functions. We would like to write these 4-point functions it terms of $G$ and $F$ so as to obtain two equations of motion for $G$ and $F$. This is usually done by approximating a pair of operators with their mean value $c c \simeq \av{cc}$, which is given by the (anomalous) Green function. By choosing a certain pair, out of the six possibilities for each 4-point function, one can obtain two coupled equations for $G$ and $F$. In Eq (\ref{G1}) we choose the approximation 
\begin{align} \label{m1}
 c^{s}_\mathbf{k+q}(\tau)  c^r_\mathbf{k'-q} (\tau')  \rightarrow \av{c^{s}_\mathbf{k+q}(\tau)  c^r_\mathbf{k'-q} (\tau')} \delta_\mathbf{k+k'}
\end{align}
and in Eq (\ref{F1}) we choose the approximation 
\begin{align}  \label{m2}
  c^{\dagger,s}_\mathbf{k-q}(\tau)  c^{\dagger,r}_\mathbf{k'}(\tau') \rightarrow \av{ c^{\dagger,s}_\mathbf{k-q}(\tau)  c^{\dagger,r}_\mathbf{k'}(\tau')} \delta_\mathbf{k+k'-q}
\end{align}
Inserting these and writing all the expectation values as Green function, according to  (\ref{G}),  (\ref{Fdag}) and  (\ref{F}) we get
\begin{align} 
(\partial_\tau + \xi_\mathbf{k})  G^{s,s'}(\mathbf{k} ,\tau) &= 
 \delta(\tau) \delta_{s,s'} - \sum_{\mathbf{q},r}  
  \int_{-\beta}^\beta d\tau' V(q,\tau-\tau') 
    F^{s,r}(\mathbf{k+q} ,\tau-\tau')  F^{\dagger,r,s'}(-\mathbf{k} ,\tau') ,\\
(\partial_\tau - \xi_\mathbf{k})  F^{\dagger,s,s'}(\mathbf{k} ,\tau) &= 
    \sum_{\mathbf{q},r}  
  \int_{-\beta}^\beta d\tau' V(q,\tau-\tau') 
   F^{\dagger,s,r}(\mathbf{k-q} ,\tau-\tau')  G^{r,s'}(\mathbf{-k} ,\tau'). 
\end{align}
The spin structure of these equations implies that $G$ is diagonal while $F$ is off diagonal so we can suppress the spin index. We now go over to Matsubara frequency and write $G(\vk,\tau) = T \sum_{\omega} e^{-i \omega \tau} G(\vk,\omega),F(\vk,\tau) =T \sum_{\omega} e^{-i \omega \tau} F(\vk,\omega),F^{\dagger}(\vk,\tau) = T\sum_{\omega} e^{-i \omega \tau} F^{\dagger}(\vk,\omega)$. We get 
\begin{align} 
(- i \omega + \xi_\mathbf{k})  G(\mathbf{k} ,\omega) &=1 - 
T \sum_{\mathbf{q},\omega'}  
    V(\vq,\omega-\omega')  F(\mathbf{k+q} ,\omega')  F^{\dagger}(\mathbf{-k} ,\omega) ,\\
(- i\omega - \xi_\mathbf{k})  F^{\dagger}(\mathbf{k} ,\omega) &= 
   T \sum_{\mathbf{q},\omega'}      V(q,\omega-\omega')
   F^{\dagger}(\mathbf{k-q} ,\omega')  G(\mathbf{-k} ,\omega) 
\end{align}
where
\begin{align}  \label{interA}
V(\vq,\omega) =  \int_{-\beta}^\beta d\tau V(\vq,\tau)  e^{i \omega \tau} =V^c_{\vq}-   \sum_\alpha { \Gamma     \abs{\phi^\alpha_\mathbf{q}}^2 \over  \left( \omega^\alpha_\mathbf{q}\right)^2 + \omega^2}. 
\end{align}
Since $\xi_\mathbf{k} = \xi_\mathbf{-k}$ and $V(\vq,\omega)=V(-\vq,\omega)$ we assume also $F^{\dagger}(\mathbf{k})=F^{\dagger}(\mathbf{-k})$ and $G(\mathbf{k} ,\omega) =G(\mathbf{-k} ,\omega) $. Now we define
\begin{align} 
\Delta(\mathbf{k} ,\omega) &= T \sum_{\mathbf{q} ,\omega'}  
 V(q,\omega-\omega')  F(\mathbf{k+q} ,\omega')  \\
 \Delta^*(\mathbf{k} ,\omega) &=- T \sum_{\mathbf{q} ,\omega'}  
 V(q,\omega-\omega')  F^{\dagger}(\mathbf{k-q} ,\omega')   \label{delt}
\end{align}
and get two algebraic equations
\begin{align} 
(-i\omega + \xi_\mathbf{k})  G(\mathbf{k} ,\omega) &=  1 -
 \Delta(\mathbf{k} ,\omega)  F^{\dagger}(\mathbf{k} ,\omega)
 ,\\
(i\omega + \xi_\mathbf{k})  F^{\dagger}(\mathbf{k} ,\omega) &= 
    \Delta^*(\mathbf{k} ,\omega)   G(\mathbf{k} ,\omega) 
\end{align}
with the solution
\begin{align} 
 F^{\dagger}(\mathbf{k} ,\omega) = { \Delta^*(\mathbf{k} ,\omega) \over \Delta^2(\mathbf{k} ,\omega) + \omega^2 + \xi^2_\mathbf{k}  } .
\end{align}
Inserting back into Eq(\ref{delt}) we have
\begin{align} 
\Delta^*(\mathbf{k} ,\omega) =-T \sum_{\mathbf{q} ,-\omega'}  
 V(\vq,-\omega-\omega') { \Delta^*(\mathbf{k+q} ,\omega) \over \Delta^2(\mathbf{k+q} ,\omega) + \omega^2 + \xi^2_\mathbf{k+q}  }   
\end{align}
which is Eq. (7) in the main text.

When employing the mean field approximation, Eq. (\ref{m1}) and (\ref{m2}), one could have chosen other pairing possibilities. These would have renormalized the dispersion relation $\xi_\vk$ and could in principle modify some parameters. The typical effect of such extension, known as Eliashberg theory or strong coupling, is to introduce a frequency dependence for the gap function, due to the dynamics of the electrons. In our case this dependence comes directly from the dynamics of the FE modes.   

 \subsection{The isotropy of the interaction at small momentum} \label{iso}

The effective interaction is given by Eq. (\ref{interA}):
\begin{align} \label{potA}
V(\vq,\omega) = - \sum_\alpha { \Gamma \abs{\phi_\mathbf{q}^\alpha}^2 \over \left( \omega^\alpha_\vq \right)^2 + \omega^2}. 
\end{align}
We omit the Coulomb interaction in this section. The frequency dependency is handled in two limits. One is when $T \gg \omega^\alpha_\vq$ and only the term $\omega = 0$ is retained. The other is is when $T \ll \omega^\alpha_\vq$ and the term $\omega^2$ is neglected. Thus the remaining expressions for the interaction, after the frequency sum, are similar in both cases:
\begin{align} 
V(\vq,\omega) \propto  \sum_\alpha {  \abs{\phi_\mathbf{q}^\alpha}^2 \over \left( \omega^\alpha_\vq \right)^2 }. 
\end{align}
The expression for $\phi_\mathbf{q}$ is given in Eq (\ref{dip}), regarding $\omega^\alpha_\vq $ we assume it has a minimum at $q=0$ and that the dispersion in the $\alpha$ direction dominates so we expand it as $\omega^\alpha_\vq \simeq \omega_0 + a q_\alpha^2$ (At the minimum the first order is likely to vanish, but in any case taking it into account would not change the result).  Inserting these, and writing explicitly the sum over $\alpha$ we have
\begin{align} \label{potB}
V(\vq,\omega)  \propto  q^{-4} \left( {q_x^2 \over (\omega_0 + a q_x^2)^2 } + {q_y^2 \over (\omega_0 + a q_y^2)^2 } +{q_z^2 \over (\omega_0 + a q_z^2)^2 } \right). 
\end{align}
For $q \rightarrow 0$, this interaction is divergent as $q^{-2(6)}$ for the case of $\omega_0 \ne (=) 0$, due to its long range coulombic nature, and, as one might expect, it becomes approximately isotropic. In any real scenario there will be some screening regulating the divergence, which can be represented, for example, by introducing the Thomas-Fermi wavevector $q_{TF}$ and replacing $q \rightarrow q + q_{TF}$. 

\subsection{The ``softening" of the longitudinal mode} \label{softLO}

As shown above, our main interest is in modes with a wave vector which is parallel to the direction of the electric dipole. These are referred to as ``longitudinal", since they couple to the electrons. We assume their frequency, for $q=0$, at the QCP, vanishes. This seems to contradict the standard treatment of ferroelectric transition, leading to the Lyddane-Sachs-Teller (LST) relation, as well as some other results \citep{millis}, which implies that only the transverse modes soften at the phase transition. As long as one consider the model that is presented in the main text, namely the quantum Ising model for electric dipoles with no interaction between different polarization, the vanishing frequency of the longitudinal modes, for $q=0$, at the QCP, is a straight forward consequence. However, if this phenomena would contradicts some general results, it would imply that the model, its application to this system or some other assumption, are not physical. In this section we show that this is not the case.   

We start with the LST relation 
\begin{align} 
{ \omega_L^2 \over \omega_T^2 } =  {\varepsilon(0) \over  \varepsilon(\infty)} 
\end{align}
where $\omega_{T(L)}$ is the frequency of the transverse (longitudinal) mode and $\varepsilon(0)$ and $ \varepsilon(\infty)$ are the static and optical dielectric constants, respectively. The standard interpretation of this equation is that at the phase transition, $\varepsilon(0)  \rightarrow \infty$ and $\omega_T \rightarrow 0$ while $\omega_L$ remains finite and does not get soft. However, one have to distinguish between two similar phenomena: (i) A ferroelectric phase transition at finite temperature and (ii) a quantum phase transition. Both cases involve a vanishing frequency. In the first, the frequency of the transverse mode vanishes as $$\omega_T \propto \sqrt{T-T_c}$$ where $T_c$ is the Curie temperature, and in the second the energy scale vanishes as $$\omega \propto (p-p_c)^{z \nu},$$ where $z \nu$ is a critical exponent and $p$ is a tunning parameter taking the value $p=p_c$ at the QCP. Our model refer to the second phenomena, where the vanishing gap have to include all modes related to the system.

The standard treatment of ferroelectric transition and the quantum criticality formalism do not contradict since they describe different regimes. When the phase transition occur at finite temperature the quantum criticality formalism implies there is an energy scale, which is larger than the critical temperature. When quantum effects dominate, for example in SrTiO3 at low temperatures, the LST relations are violated \cite{LSTviolate1,LSTviolate2}.

In \cite{millis} a quantum paraelectric-ferroelectric transition was discussed. The critical point was defined to be where the transverse modes become soft, while the longitudinal mode was considered to remain stiff. The difference with respect to our model is in the type of broken symmetry. Treating the ferroelectric transition as a breaking of a continuous symmetry, one obtain massless Goldstone modes which have to be transverse, with respect to the direction of the spontaneously chosen direction. Additional terms in the action can then gap these modes as well, for example by coupling them to other degrees of freedom. Alternatively, raising the temperature will restore the symmetry and wash out the effect of these modes. In contrast, in the Ising model only a discrete symmetry is broken so there is only one critical point where the gap vanishes, and it is the same for both longitudinal and transverse. In addition, we consider three separate Ising systems, one for each direction of the electric dipole or ``easy axis'', which do not interact. Each system reside in a 3D lattice and thus contain both longitudinal and transverse modes. They can have different critical points if the crystal is anisotropic, but they cannot gap one another, since they do not interact. 

Note that applying the assumption of no interactions between different polarization to the formalism used in \cite{millis}, which would make the action diagonal in ``polarization space'', yields indeed a soft longitudinal modes. The motivation for this assumption is due to the nature of the dipole-dipole interaction, which is much stronger when their polarization is (anti) parallel. It is similar to the reason transverse modes do not couple to itinerant electrons.            

It might be also worth to clarify the meaning of longitudinal and transverse at $q=0$. Often, the spectrum of of optical phonons is obtained by studying the response of the ions in the crystal to an external electric field and calculating the dielectric function $\varepsilon (\omega,\vq)$. The spectrum $\omega_\vq$ for different modes can then be obtained from the poles and zeros of $\varepsilon (\omega,\vq)$, which can be understood as resonances of oscillations in the crystal. Since the starting point of the derivation is a response to an external field it is natural to treat the longitudinal and transverse modes separately, even if they refer to the same ion displacements in a single unit cell. Taking the field strength to zero at the end of the calculation does not matter as for any finite $\vq$ the dispersion of the modes is different. Considering the crystal as an independent system, this distinction is not well defined at $q=0$ since there is no preferred direction. Typically this is not a problem as the impact of a single point in momentum space on any physics can be neglected.

In the quantum criticality formalism, one typically calculates the spectrum of some Hamiltonian. If the degrees of freedom in the Hamiltonian refer to electric polarization, they should carry an additional index, denoting the orientation of the polarization. The direction of the wave vector, with respect to that orientation, distinguish between longitudinal and transverse modes, but at $q=0$ this distinction disappear. In our system the dispersion is assumed to have a minimum at $q=0$ and since at a quantum phase transition a gap has to close, the frequency at this point have to vanish.  

The problem then boils down to the question how big is the volume in momentum space where the frequency of the longitudinal mode is small, or in other words what is its dispersion at small $q$. In this work we consider a delta function in $q$, $V(\mathbf{q}) \simeq - g \delta(q)$ so the issue is incorporated in the magnitude of coupling constant $g$. A small volume in momentum space would mean small $g$ and thus a small superconducting temperature $T_c$. It is important to remember that the context here is the low doping regime. In this case the dispersion of the electrons is weak, so a small volume might suffice and $T_c$ is small. Thus it is possible that the pairing at this scenario is due to this region in momentum space. 
    
%


  \end{widetext}
  
\end{document}